\documentclass[prl,twocolumn,showpacs,preprintnumbers,amsmath,amssymb]{revtex4}

\usepackage{graphicx}
\usepackage{dcolumn}

\newcommand{\beq}{\begin{equation}}
\newcommand{\eeq}{\end{equation}}
\newcommand{\beqa}{\begin{eqnarray}}
\newcommand{\eeqa}{\end{eqnarray}}

\begin{document}
\title{Theory of isotope effect in photoemission spectra of 
high-T$_c$ superconducting cuprates}

\author{G. Seibold$^{1}$ and M. Grilli$^2$} 

\affiliation{$^1$Institut f\"ur Physik, BTU Cottbus, PBox 101344,
         03013 Cottbus, Germany}
\affiliation{$^2$Istituto Nazionale di Fisica della Materia, 
Unit\`a Roma 1 and SMC Center, and Dipartimento di Fisica\\
Universit\`a di Roma "La Sapienza" piazzale Aldo Moro 5, I-00185 Roma, Italy}

\begin{abstract}
We investigate the effect of isotope substitution on the electronic
spectral functions within a model where the charge carriers are
coupled to bosonic charge-order (CO) fluctuations centered around some 
mean frequency $\omega_0$ and with enhanced scattering at
wave-vector q$_c$. It is shown that a mass dependence of $\omega_0$ is not
sufficient in order to account, especially at high energies, for the 
dispersion shifts experimentally observed in an optimally doped 
superconducting cuprate. We argue that isotope substitution
induces a change of the spatial CO correlations which gives good
account of the experimental data.
\end{abstract}
\date{\today}
\maketitle

Several isotope effects (IE) have been found in underdoped
superconducting cuprates, not only  in the critical temperature
but also in the superfluid density,  in the fermion effective mass,
and in the dynamic lattice distortion \cite{KELLER} and
in the pseudogap crossover temperature $T^*$ \cite{RUBIO_TEMPRANO}.
More recently a remarkable IE has also been found in angle-resolved
photoemission spectroscopy (ARPES) experiments on optimally doped Bi-2212 \cite{GWEON}.
In this experiment the dispersive single-particle
excitations were investigated and strong peculiar variations upon
replacing $^{16}O$ with  $^{18}O$ were found. Specifically,
little isotopic dependence was found both for the kink and the
low-energy part of the spectra along the $\Gamma \to X$,
[i.e., $(0,0)\to (\pi, \pi)$] direction. On the other hand the dispersive 
``hump'' at higher binding energy displayed a sizable IE,
which caused a shift of about 15 meV towards larger binding energies.
Around the M points at $(\pi,0)$ and $(0,\pi)$ the IE {\it has a reversed sign} 
and both the low- and high-energy parts of the dispersion shift to 
smaller binding energies upon isotope substitution. 
In this region shifts as large as 40 meV are reported.
This IE is absolutely non trivial in various respects. First of all,
the ion replacement is expected to change by  only 6\% the phonon frequencies
involving oxygen ions. The changes of the electron dispersions are
large and  within a standard perturbative 
Eliashberg-like scheme would imply too large e-ph
couplings \cite{maksimov}. Secondly the isotopic shifts
have different signs depending on the region of the Fermi surface.
Thirdly they do not substantially occur around the ``kinks''  of the
dispersion curves, as it would be expected if the isotope replacement
were only changing the frequency of the phonon [or more generally of the 
collective mode (CM)] interacting with the quasiparticles (QP's).

In this letter we specifically address the issue of 
this peculiar IE in ARPES and we provide a natural explanation for it within
the theory based on the occurrence of a charge-ordering (CO)
quantum-critical point (QCP) around optimal doping of the 
cuprates \cite{CDG}. This theory predicts 
the occurrence of a CO phase in a
substantial region of the phase diagram below a doping-dependent critical
temperature $T_{CO} (x)$ ending at zero temperature in a QCP around
optimal doping. Actually several mechanisms (pairing, disorder, and low
dimensionality) may prevent a fully realized long-range
CO state. Nevertheless, despite this ``missed'' criticality,
in the low-temperature  underdoped region
below $T^*\sim T_{CO}$, local dynamical CO
should be revealed as a natural inclination of the system.
 After this theoretical prediction, many evidences
have arisen that a QCP could be hidden under the superconducting ``dome''
\cite{OLDQCP,NEWQCP}. The prediction of a local 
CO in the underdoped cuprates is also finding new experimental confirmations
\cite{exptCO}, which corroborate less recent observations of ``stripe''
phases. Within the CO-QCP theory it was also possible to explain
the peculiar IE of the pseudogap crossover temperature $T^*$  mentioned
above \cite{ANDERGASSEN}.
Therefore it seems legitimate to adopt this theoretical framework
to address the issue of IE in ARPES experiments. In particular there
are several characteristics of the theory, which render it a particularly
suitable scheme: a) the presence of a ``missed'' criticality
justifies the presence of low-energy (nearly critical) charge collective
modes (CM), which have already been shown to produce ``kinks'' in the
dispersive spectra \cite{SG}; b) on general grounds, the proximity to
a (``missed'') criticality renders the system highly susceptible so that
large effects (e.g. large IE on the single-particle excitations) can
arise from small changes (like a 6\% shift of some phonon frequencies);
c) within the Hubbard-Holstein model, which was the starting point of the
CO-QCP theory, the CM are mixed phonon-plasmon modes, which therefore have 
both a built-in dependence on the phonon frequency naturally leading to IE's
and an electronic character, which renders them similar to the spin modes
considered in Refs.\cite{ESCHRIG};
d) the CM's are only critical near specific wavevectors ${\bf q}_c$ related to
the wavelengths of the CO textures. As a consequence they have a
strong momentum dependence, which allows them to play different roles in interacting
with QP's in different regions of the Brillouin zone. As we shall see
this is a crucial feature in order to explain the opposite sign of the IE
on dispersive bands along the nodal and antinodal directions.
Within the above framework we performed a systematic analysis of the
effects of changing the CM frequency $\omega_0$, the QP-CM coupling 
$g$, and the CO inverse correlation length $\gamma$.
Our main finding is that changes of $\omega_0$ and/or $g$ alone by no means
allow to reproduce the experimental results. On the other hand the
crucial parameter to vary in order to semiquantitatively reproduce the
experimental isotopic shift is the inverse charge-charge coherence length $\gamma$.
Before reporting on our detailed calculations,
we first present qualitative arguments to give an intuitive understanding of
this result. We start, e.g, from an hole state at wavevector ${\bf p}$, which is scattered
by a narrow CM of energy $\omega_0$ and wavevector ${\bf q_c}$. This hole acquires a
self-energy $\Sigma_p(\omega)=g^2/ (\omega+\omega_0-\varepsilon_{\bf p+q_c})$.
If the CM has instead a wavevector broadening $\gamma$, the 
self-energy becomes proportional to the hole density of states (DOS) 
$\rho_{\bf p+q_c}$ in the
region of the momenta, where the hole is scattered by the CM.
In the two-dimensional case of our interest this region
has an area $\gamma^2$ around ${\bf p+q_c}$.
If the DOS does not vary appreciably in the Brillouin zone, one obtains
a result similar to the one found by Engelsberg-Schrieffer 
\cite{ENGELSBERG} which was derived under the
assumption of electron-hole symmetry and a q-independent
matrix element between the electrons and the CM (a Holstein phonon):
There is a dispersion break at a binding energy $\omega \sim \omega_0$
and a reduction of $\omega_0$ always leads to a shift of the
electron dispersion at higher binding energies (and only in a limited 
energy range around the break). On the other hand, if 
$\rho_{\bf p+q_c}$ varies substantially  around the Fermi surface,
a different situation can occur. Let us suppose that
the isotope replacement
acts on $\gamma$ [e.g., $\gamma(^{18}O) > \gamma(^{16}O)$,
this is what we find to be the relevant effect in our detailed
calculations]. Then upon isotopic substitution the
hole can be scattered in a broader momentum region where the average
DOS can be substantially different. In particular, if the local DOS
increases 
$Re \Sigma(^{16}O,\omega<-\omega_{0}) < Re \Sigma(^{18}O,\omega<-\omega_{0})$ 
(cf., e.g., Fig. \ref{self}c) and as a result the 
 electronic dispersion is shifted upwards to
lower binding energies
$\varepsilon_k(^{18}O) > \varepsilon_k(^{16}O)$. 
This is indeed the desired effect to explain
the experiments of Ref. \cite{GWEON} near the M points.

In order to substantiate these ideas within more detailed calculations 
we consider a system of superconducting electrons exposed to
an effective action
\begin{equation}
S=-g^2 \sum_{\bf q}  \int_0^{\beta}d\tau_1\int_0^{\beta}d\tau_2
\chi_{\bf q}(\tau_1-\tau_2) \rho_{\bf q}(\tau_1) \rho_{-{\bf q}}(\tau_2)
\end{equation}
describing dynamical incommensurate CO fluctuations.
Using Nambu-Gorkov notation the unperturbed Matsubara Green's functions
$\underline{\underline{G}}^0$
are $2\times 2$ matrices and
the leading order one-loop contribution to the self-energy reads
as
\begin{equation}\label{SELF}
\underline{\underline{\Sigma}}({\bf k},i\omega)=-\frac{g^2}{\beta}
\sum_{{\bf q},ip} \chi_{\bf q}(ip)
\underline{\underline{\tau_z G}}^0({\bf k}-{\bf q},i\omega-ip)
\underline{\underline{\tau_z}}
\end{equation}
which in turn allows for the calculation of $
\underline{\underline{G}}=\underline{\underline{G}}^0+
\underline{\underline{G}}^0 \underline{\underline{\Sigma}}
\underline{\underline{G}}$.
The spectral function can be extracted from 
$A_{\bf k}(\omega)=(1/\pi)Im G_{11}({\bf k},\omega)$.

In order to simplify the calculations we consider a Kampf-Schrieffer-type
model susceptibility \cite{KAMPF}
which is factorized into an $\omega$- and q-dependent part, i.e.
\begin{equation}\label{chi}
\chi_{\bf q}(i\omega)=W(i\omega)J({\bf q})
\end{equation}
where, choosing a unit lattice spacing, 
\begin{equation}\label{jq}
J({\bf q})=\frac{{\cal N}}{4}
\sum\limits_{\pm q_x^c;\pm q_y^c}\frac{\gamma}{\gamma^2
+2-\cos(q_x-q_x^c)-\cos(q_y-q_y^c)}.
\end{equation}
${\cal N}$ is a suitable normalization factor introduced to
keep the total scattering strength constant while varying $\gamma$.
The (real-frequency)-dependent part 
$W(\omega)=\int d\nu F_{\omega_0,\Gamma}(\nu) 2\nu/(\omega^2-\nu^2)$
with $F_{\omega_0,\Gamma}(\omega)\sim \Gamma/[(\omega-\omega_0)^2 +\Gamma^2]$
is a normalized
lorentzian distribution function centered around $\omega_0$ with 
halfwidth $\Gamma$. If only this part were present in $\chi(i\omega)$,
one would have a bosonic spectrum
$B(\omega)= \mbox{tanh}(\omega/(kT)) F_{\omega_0,\Gamma}(\omega)$
which is a ``smeared'' version of the Holstein phonon considered
in Ref.\cite{ENGELSBERG}. The crucial feature of the susceptibility
(\ref{chi}) is the substantial momentum dependence, which describes
the local order formation and reflects the proximity to a (missed) CO
instability. This susceptibility 
contains the charge-charge correlations which are enhanced at the four
equivalent critical wave vectors $(\pm \pi/2,0)$, $(0,\pm \pi/2)$.
The static limit $g(\nu)=\delta(\nu)$  together with an 
infinite charge-charge correlation length $\gamma \rightarrow 0$
reproduces mean-field results for a long-range CO phase \cite{GOETZ}.
For the bare electron dispersion used in the following
\begin{displaymath}
\varepsilon_k = -2t(\cos(k_x)+\cos(k_y)) + 4t' \cos(k_x)\cos(k_y) -u
\end{displaymath}
the parameters have been fitted so that the resulting spectral functions
approximately agree in energy with the data of Ref. \cite{GWEON}.
We use $t=0.35 eV$, $t'/t = 0.32$, $u=-0.42eV$.
The superconducting d-wave gap is $\Delta_k = 35 meV [\cos(k_x)-\cos(k_y)]/2$.
The width of the lorentzian in frequency space is $\Gamma = 50 meV$

In Ref. \cite{GOETZ} we demonstrated that a static
two-dimensional eggbox-type charge modulation with ${\bf q_c}$
oriented along the (1,0), (0,1) directions can account for the basic
Fermi surface (FS) features in the optimally and underdoped Bi2212 compounds.
In a subsequent paper \cite{SG} we analyzed ARPES spectra
in Bi-2212 showing that the main dispersive features \cite{CAMPU,KAMINSKI}
were reproduced, including the well-known kink in the nodal dispersion
which appears as a break for momentum scans far from the node 
\cite{KAMINSKI}. The main point to be recollected here
is that the M-points define the ``hot'' regions for incommensurate CO
scattering. In these regions of k-space the low-energy QP's 
are strongly scattered by the CM's into other states with 
high DOS (still near the M points).
On the other hand, along $\Gamma -X$ the coupling of electronic states
to the CO mode is significantly reduced (``cold'' region) at low
binding energies and one finds weakly interacting QP's
which get heavily damped only at high binding energies.
Here the QP's both those at low and at high binding energies
are scattered by the CM's into regions with rather low DOS.
This different character between ``hot'' and the ``cold'' regions
is a key point to understand the 
behavior of the ARPES-IE, with the ``surprising'' change of sign
moving from diagonal scans to scans near M and sets the stage for
our analysis. 

{\it --- Only mode-frequency change ---}
First of all we tried a simple change of the CM frequency only.
Since we are not dealing with simple phonons, but rather with
(nearly critical) CM's, it is difficult to {\it a priori}
determine the  isotopic shift (if any) of the CM frequency. We take
 $\omega_0(^{16}O)=75 meV$, $\omega_0(^{18}O)=60 meV$
to optimize the agreement with experiments.
The frequency shift is quite large ($15 meV$) with respect to
the one expected for phonons.
The coupling constant is $g(^{16}O)=g(^{18}O)=0.13 eV$,
which leads to self-energies at the nodal point 
of the same order of magnitude than those reported in \cite{GWEON}.
Both along the diagonal direction and along the vertical cut 
corresponding to cut $\#7$ in Ref. \cite{GWEON} 
there is a significant isotope shift only 
in a narrow energy window near the kink around $-0.1 eV$.
Our dispersions do not display an isotope
shift in the high energy region.
Upon changing in the opposite way the frequency
 $\omega_0(^{16}O)<\omega_0(^{18}O)$
the agreement with experiments is even worse.
Therefore, within our model and despite the large shift in 
$\omega_0$ the IE can not be ascribed to a simple shift
of the CM frequency.

{\it --- Coupling and mode-frequency change ---}
As before we choose 
$\omega_0(^{16}O)=75 meV$ and $\omega_0(^{18}O)=60 meV$.
In addition we consider now different coupling constants for the isotopes. 
$g$ can be determined from the requirement that the
normal contribution to the self-energy $\Sigma_{11}(k_F,E_F)$ at the
nodal point gives a reasonable fit to the $Re \Sigma$ extracted in
Ref. \cite{GWEON}.
We find $g(^{16}O)=0.13 eV$ and $g(^{18}O)=0.114 eV$. 
Along the diagonal direction one finds qualitatively the same behavior
as with $\omega_0$ changes only: the IE is still confined to a rather 
narrow energy window. Moreover the theoretical dispersions along the vertical cut
do not properly describe the findings of Ref.\cite{GWEON}: 
contrary to the experiments, with isotopic replacement
the low-energy part of the dispersion for $^{18}O$ 
is shifted at higher energies with respect to 
the dispersion for $^{16}O$. Furthermore, while $^{16}O$ shows
the break in the dispersion as in the experiment, 
the lower coupling for $^{18}O$ turns this break into a kink. 

\begin{figure}[hbt]
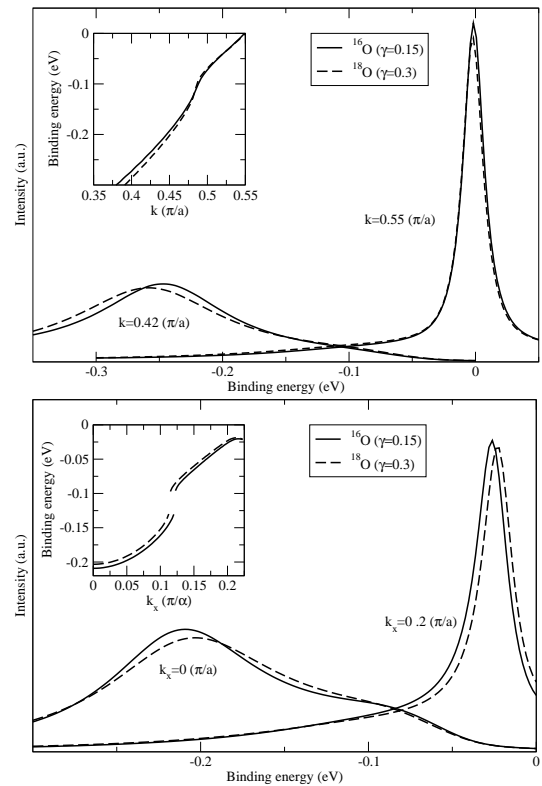

\includegraphics[width=7cm,clip=true]{fig1a.eps}
\includegraphics[width=7cm,clip=true]{fig1b.eps}
\caption{EDC's and the corresponding 
EDC-derived dispersions (inset) for the case where only $\gamma$ is varied upon
isotope substitution.
Intensities have {\it not} been scaled to the same amplitudes as
done in Ref. \protect\cite{GWEON}. 
Upper panel: diagonal cut;
Lower panel: vertical cut at $k_y=0.6 (\pi/a)$ corresponding
to cut 7 in Fig. 2 of Ref. \protect\cite{GWEON} 
(cf. inset to Fig. \protect\ref{self}d).
}
\label{fig1}
\end{figure}

\begin{figure}[hbt]
\includegraphics[width=8cm,clip=true]{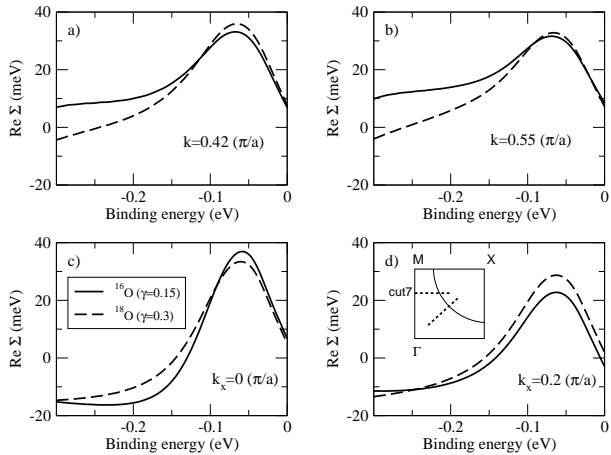}
\caption{Real part of the normal contribution to the self-energy
$Re \Sigma_{11}(q,\omega)$. Panels (a,b) show  $Re \Sigma_{11}$
for two momenta along the nodal direction and panels (b,c)
are for two momenta on cut $\#7$ (cf. inset to d). }
\label{self}
\end{figure}

{\it --- Only change of $\gamma$ ---}
Finally we consider a change of $\gamma$ alone.
We take $\omega_0(^{16}O)=\omega_0(^{18}O)=50 meV$ 
and $g(^{16}O)=g(^{18}O=0.16 eV$, but we
choose different spatial correlation lengths,
i.e. $\gamma=0.15$ for $^{16}O$ and $\gamma=0.30$ for $^{18}O$.
Fig. \ref{fig1} shows energy distribution curves (EDC's) and the corresponding
dispersions for a cut along the diagonal (Fig. \ref{fig1} upper panel) and
a vertical cut near the M-point similar to cut $\#7$ in Ref. \cite{GWEON}
(Fig. \ref{fig1} lower panel, cf. inset to Fig. \ref{self}d).
Along the diagonal direction, below the kink energy we find 
a  shift of the $^{18}O$ dispersion towards larger binding energies.
Similarly to what is experimentally observed, the IE shift extends
far below the kink energy. In the low energy regime above
the kink the shift becomes negligible like in the experiments.
One only finds  a minor discrepancy with experiment
since in the crossover region the theoretical $^{16}O$ dispersion falls below
the $^{18}O$ one. 
In the vertical cut as shown in Fig. \ref{fig1} (lower panel) 
one sees that the $^{18}O$ dispersion is shifted
{\it above} the $^{16}O$ dispersion both at low and high binding energies.
This behavior is also in agreement with the experimental findings
of Ref. \cite{GWEON}.
We notice that, despite the qualitative agreement, the calculated
shift ($\sim 15-20 meV$) is substantially smaller than the experimental one
($\sim 30-40 meV$).  However, this quantitative discrepancy could be
solved by assuming a (more realistic) momentum dependence of the QP-CM 
mode coupling $g$, which instead we take constant for simplicity
reasons.

Fig. \ref{self} shows $Re \Sigma_{11}(q,\omega)$ for the 
case where $\gamma$ is increased by isotope substitution, while keeping $\omega_0$
and $g$ fixed.
Results for two momenta along diagonal cuts are reported in the two top 
panels, where
the most prominent feature is that at large binding energies 
the self-energy generically 
decreases upon increasing $\gamma$ (i.e. upon decreasing the CO correlation
length).
Remarkably this reduction persists down to very high binding energies.
While this reduction is expected on general grounds even for the phonon case,
\cite{maksimov}, the frequency extension of this effect is rather unusual
and arises from the plateau in the high-energy part of the self-energy 
occurring when $\gamma$ is small.
On the other hand the self-energy generically varies little at low energies,
while a moderate increase, $Re \Sigma(^{18}O) > Re \Sigma(^{16}O)$, 
 occurs near its maximum (around which the kink
in the dispersion takes place) due to the enhancement of the phase space 
of the scattered QP states induced by the increased $\gamma$ 
\cite{notemaksimov}. 
This effect is even more pronounced in the antinodal region 
reported in the two lower panels. Along cut $\#7$ one therefore observes 
a shift of the $^{18}O$ dispersion to smaller binding energies.

{\it --- Conclusions ---}
We applied here a simple perturbative scheme of superconducting
quasiparticles coupled to charge CM's.
From the above systematic analysis we find that, if the isotopic
substitution changes the CM momentum broadening $\gamma$, our model
accounts both for the large energy range in which
 isotopic shifts are observed and for the change in sign of these
shifts moving from the $\Gamma \to X$ cuts to the cuts near the M points.
The fact that the crucial effect of the isotopic replacement is to
change the CO coherence length (inversely related to $\gamma$)
also supports the idea that charge fluctuations are highly
correlated in optimally doped cuprates. This is consistent with 
Ref. \cite{ANDERGASSEN}, where it was found that the $^{16}O \to ^{18}O$ 
replacement shifts the (``missed'') quantum-critical line to
higher dopings (or equivalently confers to the system
properties of more underdoped materials). For  optimally
or underdoped materials at low temperature this ``pushes'' the system more
deeply in the nearly ordered region away from the
``missed'' critical line, naturally reducing the
spatial CO correlations. An additional broadening of
the CM momentum distribution can arise from disorder, if
the $^{16}O \to {^{18}O}$ substitution is incomplete.
Another possible way of arguing can be derived from Gutzwiller+RPA 
calculations of the stripe phase, which indicate that upon underdoping the
intensity distribution of the mode becomes more extended 
in momentum space \cite{SEIBOLD}.
All these arguments provide support to an increase of $\gamma$ upon
$^{16}O \to ^{18}O$ substitution.

Since it turns out that isotopic changes of $\omega_0$ and of $g$
play a minor role, one could argue that the CM's not necessarily are
charge modes (mixed with phonons), but could as well be spin modes.
Indeed, our perturbative scheme previously adopted in Ref. \cite{SG}
provided an interpretation
of spectral features in Bi-2212 quite similar to those of Ref. 
\cite{ESCHRIG}, where the coupling of electrons to antiferromagnetic
(AF) spin CM's was considered.
This is natural because for both AF and CO scattering the 
states near the $M$ points are the ``hot spots'', i.e. are most strongly
affected by the scattering. To (at least preliminarly) discriminate between
charge and spin modes, we also checked 
whether CM's peaked around ${\bf Q}_{AF}=(\pi,\pi)$ relevant for AF
fluctuations could account for
the IE in ARPES. While a more detailed analysis of AF modes
is beyond the scope of this work, we found that CM's peaked at ${\bf Q}_{AF}$
fail in reproducing the major features of the experimental 
isotopic shifts in the dispersion curves. 
Although a more specific analysis for AF spin modes should be carried out,
our analysis suggests that the critical momenta typical of charge modes
[stripes and ``eggbox'', with ${\bf q}_c\sim (\pm \pi/2,0),\, (0,\pm \pi/2)$] 
are an important ingredient to account for ARPES experimental findings.

We acknowledge interesting discussions with C. Castellani, C. Di Castro, 
G.-H. Gweon, A. Lanzara, who we also thank for showing us her data before publication, and
J.Lorenzana. M. G.  acknowledges financial support from MIUR Cofin 2003
and G. S. from the Deutsche Forschungsgemeinschaft.


\begin{thebibliography}{99}
\bibitem{KELLER} For a review see, e.g.,
H. Keller, {\it Unconventional isotope effects in cuprate superconductors},
in {\it Structure and Bonding}, Springer Verlag, 2004, and references therein.
\bibitem{RUBIO_TEMPRANO} D. Rubio Temprano, Phys. Rev. Lett. {\bf 84}, 1990 (2000).
\bibitem{GWEON} G.-H. Gweon, {\it et al.}, Nature {\bf 430}, 187 (2004).
\bibitem{maksimov} E. G. Maksimov, {\it et al.}, cond-mat/0408251.
\bibitem{CDG} C. Castellani, C. Di Castro, and M. Grilli,
Phys. Rev. Lett. {\bf 75}, 4650 (1995).
\bibitem{OLDQCP} For a list of early evidences 
see C. Castellani, {\it et al.}, J. of Phys. and Chem. of Sol. {\bf 59}, 1694 (1998);
J.L. Tallon, J.W. Loram, Physica C {\bf 349}, 53 (2001).
\bibitem{NEWQCP} C. M. Varma, cond-mat/0312385 and references therein;
Y. Dagan, {\it et al.}, Phys. Rev. Lett. {\bf 92}, 167001 (2004);
Y. Ando, {\it et al., ibid.}, 247004 (2004);
L. Alff, {\it et al.}, Nature {\bf 422}, 698 (2003).
\bibitem{exptCO}C. Howald, {\it et al.}, Phys. Rev. B {\bf 67}, 014533 (2003);
M. Vershihin, {\it et al.}, Science {\bf 303}, 1995 (2004);
McElroy, {\it et al.}, cond-mat/0406491.
\bibitem{ANDERGASSEN} S. Andergassen, {\it et al.}, 
Phys. Rev. Lett. {\bf 87}, 056401 (2001).
\bibitem{SG} G. Seibold and M. Grilli, Phys. Rev. B {\bf 63} 224505 (2001).
\bibitem{ESCHRIG} M. Eschrig and M. R. Norman, Phys. Rev. Lett. {\bf 85}, 3261 (2000),
A. V. Chubukov and M. R. Norman, cond-mat/0402304.
\bibitem{ENGELSBERG}S. Engelsberg and J. R. Schrieffer, Phys. Rev. {\bf 131},
        993 (1963).
\bibitem{lanz01} A. Lanzara, {\it et al.}, Nature {\bf 412}, 510 (2001).
\bibitem{gromko03} A. Gromko {\it et al.}, Phys. Rev. B {\bf 68}, 174520 (2003).
\bibitem{gweon04} G.-H. Gweon, {\it et al.},J. Phys. Chem. Solids {\bf 65}, 1397 (2004).
\bibitem{KAMPF} A. P. Kampf and J. R. Schrieffer, Phys. Rev. B{\bf 42}, 7967 (1990).
\bibitem{GOETZ}G. Seibold, {\it et al.}, Eur. Phys. J. B {\bf 13}, 87 (2000).
\bibitem{CAMPU} J. C. Campuzano, {\it et al.}, Phys. Rev. Lett. {\bf 83}, 3709 (1999).
\bibitem{KAMINSKI}A. Kaminski, {\it et al.}, Phys. Rev. Lett. {\bf 86}, 1070 (2001).
\bibitem{NORMAN3} M. R. Norman, Phys. Rev. B{\bf 61}, 16117 (2000).
\bibitem{notemaksimov}
This finding differs from the result of Ref. \cite{maksimov}, where 
calculations were carried out for an Holstein phonon 
within a particle-hole symmetric model and
keeping (within our notations) the $g^2/\omega_0$ independent on the
isotopic mass.
\bibitem{SEIBOLD}G. Seibold and M. Grilli, to be published.
\end{thebibliography}
\end{document}